\documentclass[aps,twocolumn,prd,superscriptaddress,showpacs,10pt,longbibliography,footinbib]{revtex4-2}

\usepackage{graphicx}
\usepackage{dcolumn}
\usepackage{bm}

\usepackage[utf8]{inputenc}
\usepackage[T1]{fontenc}
\usepackage{mathptmx}
\usepackage{multirow}
\usepackage{changes}
\usepackage{enumitem}
\usepackage{float}
\usepackage{amsmath,amssymb,wasysym,units,pifont,footnote,xcolor,lineno,upgreek}

\usepackage[hidelinks,colorlinks=true,linkcolor=blue,citecolor=blue]{hyperref}

\begin{document}

\preprint{AIP/123-QED}

\title{Low thermal noise mirror coatings utilising titanium dioxide and germanium dioxide mixtures}

\author{M. Fazio}
 \email{mariana.fazio@strath..ac.uk}
\affiliation{SUPA, University of Strathclyde, Glasgow, UK.}
\author{I. W. Martin}
 \email{iain.martin@glasgow.ac.uk}
\affiliation{SUPA, Institute for Gravitational Research, University of Glasgow, Glasgow, UK.}
\author{P. Hill}%
\affiliation{SUPA, University of Strathclyde, Glasgow, UK.}
\author{M. Ben Yaala}
\affiliation{SUPA, University of Strathclyde, Glasgow, UK.}
\author{M. Chicoine}
\affiliation{Université de Montréal, Montréal, Québec, Canada.}
\author{C. Clark}%
\affiliation{Helia Photonics Ltd., Livingston, UK.}
\author{N. Demos}%
\affiliation{Massachusetts Institute of Technology, Cambridge, Massachusetts, USA}
\author{M. M. Fejer}%
\affiliation{Stanford University, Stanford, CA, USA}
\author{D. Gibson}
\affiliation{SUPA, University of the West of Scotland, Paisley, UK.}
\author{S. Gras}%
\affiliation{Massachusetts Institute of Technology, Cambridge, Massachusetts, USA}
\author{J. Hough}
\affiliation{SUPA, Institute for Gravitational Research, University of Glasgow, Glasgow, UK.}
\author{A. Markosyan}%
\affiliation{Stanford University, Stanford, CA, USA}
\author{G. McGhee}%
\affiliation{SUPA, Institute for Gravitational Research, University of Glasgow, Glasgow, UK.}
\author{S. Rowan}%
\affiliation{SUPA, Institute for Gravitational Research, University of Glasgow, Glasgow, UK.}
\author{J. Smith}%
\affiliation{California State University, Fullerton, USA}
\author{F. Schiettekatte}
\affiliation{Université de Montréal, Montréal, Québec, Canada.}
\author{S. Tait}
\affiliation{SUPA, Institute for Gravitational Research, University of Glasgow, Glasgow, UK.}
\affiliation{California Institute of Technology, Pasadena, California, USA}
\author{G. Vajente}%
\affiliation{California Institute of Technology, Pasadena, California, USA}
\author{S. Reid}
\affiliation{SUPA, University of Strathclyde, Glasgow, UK.}


\date{\today}

\begin{abstract}
Upgrades to ground-based gravitational-wave observatories will require mirror coatings with reduced thermal noise, enabling improved detector sensitivity and extended  astrophysical reach. Recent studies have shown that optical coatings utilising amorphous materials that exhibit a larger fraction of corner-sharing between adjacent structural units of metal-centered polyhedra are a promising route for reducing mechanical dissipation and thus thermal noise at room temperature. We report on multilayer optical coatings that are fabricated using germanium dioxide mixed with titanium dioxide (TiO$_2$:GeO$_2$) for the high index layers, and silicon dioxide (SiO$_2$) for the low index material. Single layers of TiO$_2$:GeO$_2$ are characterised to optimise the mixture proportion and based on that highly reflective multilayer stacks were deposited. Exceptional optical absorption at 1064 nm below 1 part-per-million (ppm) is observed in the multilayer stacks after heat treatment. The annealing process also induces the formation of blisters which leads to increased optical scattering. However, there is indication that blisters can be suppressed by decreasing the water partial pressure in the deposition chamber. Direct thermal noise measurements provide experimental verification of a significant 25\% reduction of thermal noise over the mirrors currently employed, which combined with sub-ppm levels of optical absorption show the potential of TiO$_2$:GeO$_2$ to improve the sensitivity of gravitational-wave observatories.

\end{abstract}

\maketitle

\section{Introduction}

Since the first detection of gravitational waves in 2015~\cite{Detection_paper}, over 100 astrophysical signals have been detected by the Advanced LIGO~\cite{Aasi_2015} and Advanced Virgo~\cite{Acernese_2015} gravitational-wave detectors. The majority of these signals arose from merging black holes, however merging neutron stars and black hole neutron star mergers have also been observed~\cite{catalogueO3a,gwtc3}. 

Gravitational waves exert transverse quadrupole strains on space, and can be detected by using laser interferometry to monitor the relative separation of test-mass mirrors at the ends of kilometer-scale perpendicular arms. Coating thermal noise, arising from thermally-induced vibrations in the detector mirrors, is a major limit to detector sensitivity. The development of improved coatings is therefore essential to enable future detectors and detector upgrades to meet their sensitivity goals, providing a significant increase in the number of detections and leading to the detection of new types of astrophysical sources.

The amplitude spectral density of coating thermal noise, $x(f)$, as measured by a laser beam of radius $w$ incident on the coating, can be written in simplified form as~\cite{harry2002thermal}
\begin{equation}
    x(f) \propto \sqrt{\frac{k_{\rm B}Td}{f w^2}\phi}
\end{equation}
where $k_{\rm B}$ is the Boltzmann constant, $T$ is the temperature, $d$ is the coating thickness, $f$ is frequency and $\phi$ is the mechanical loss of the coating. We assume for simplicity that the mechanical losses associated with bulk motion and shear motion~\cite{Hong2013} are approximately equal.
Current GW detectors use coatings composed of alternating layers of silica (SiO$_2$) low-index layers and titania-doped tantala (TiO$_2$:Ta$_2$O$_5$) high-index layers~\cite{Granata_2020}. The mechanical loss, and  therefore the thermal noise, is dominated by the TiO$_2$:Ta$_2$O$_5$ layers~\cite{Harry2006,Harry_2007,Flaminio2010,Amato_2021}, and thus an alternative high-index material is required. 

One candidate material that has been identified is germania mixed with titania (TiO$_2$:GeO$_2$)~\cite{vajente2021low}. Mechanical loss measurements of single layers composed of 44\% TiO$_2$ and 56\% GeO$_2$ have shown mechanical loss as low as 1$\times10^{-4}$. This is in good agreement with recent studies that show a link between the decrease in the ratio of edge-sharing to corner-sharing metal-metal polyhedra and a reduction in mechanical loss ~\cite{prasai2019high}.

When used alongside silica low-index layers, this has the potential to provide a significant reduction in coating thermal noise, potentially meeting the requirements of the A+ LIGO detector. However, these results are based on single-layers of TiO$_2$:GeO$_2$ and have not been confirmed for highly reflective (HR) stacks. In this letter, we present the optical and structural characterisation of TiO$_2$:GeO$_2$ single layers with various dopant concentrations after heat treatment at different temperatures and show the expected low mechanical loss of this material. Based on this, a TiO$_2$:GeO$_2$/SiO$_2$ HR stack was designed and deposited. This HR stack is shown to feature a coating thermal noise reduction of 25\% compared to the current end test mass coatings of Advanced LIGO and Advanced Virgo and sub-ppm optical absorption. Blisters are observed to form after heat treatment that increased scattering losses, however there are indications that these can be suppressed by reducing the water partial pressure in the deposition chamber.

\section{Experimental}

The coatings studied here were deposited by Cutting Edge Coatings \footnote{Cutting Edge Coatings GmbH, Hollerithallee 18, 30419 Hannover, Germany}, using ion beam deposition (IBD) with Ar as the sputtering gas~\cite{sakiew2020large}.

To allow investigation of the effect of composition on the coating properties, a number of single layers of TiO$_2$:GeO$_2$ were deposited with differing concentrations of TiO$_2$ doping. In addition, an HR multilayer stack was deposited, using 27 pairs of SiO$_2$ and TiO$_2$:GeO$_2$ layers to give a reflectivity of $\approx$99.9995\%. A cation ratio of 0.429 was selected for the TiO$_2$:GeO$_2$ layers, as a compromise between high refractive index whilst not increasing the titanium content significantly, which would unfavourably lower the crystallisation temperature. The previous studies on single-layer TiO$_2$:GeO$_2$ films ~\cite{vajente2021low} also informed target composition.

The composition of the coatings was measured using Rutherford backscattering spectrometry (RBS)~\cite{Backscattering_spectrometry}. These measurements were conducted with a 2032\,keV He beam on a Tandetron accelerator, with an incidence angle of $7^{\circ}$ and the detector placed at a scattering angle of $170^{\circ}$. Simulations with the ion beam analysis software SIMNRA~\cite{mayer1999simnra}, using a uniform  layer model, were used to estimate the element concentrations in the coatings.

The optical absorption of the single layer coatings was measured using photothermal common-path interferometry~\cite{Alexei_PCI}. In this technique, a relatively intense `pump' laser beam at the wavelength of interest is passed through a sample. Localised heating due to optical absorption of this beam creates a thermal lens in the sample. A larger, less intense `probe' laser beam is used to measure the strength of the thermal lens. The part of the probe beam which passes through the thermal lens experiences a phase change, and interferes with the unaffected part of the probe beam. The interference pattern is imaged onto a photodetector, and the absorption found by comparison with the signal from a calibration sample of well-known absorption.

The optical constants of the single layers were obtained from transmittance measurements performed in a Photon RT spectrophotometer. The wavelength was varied between 185 nm and 2000 nm and measurements were fitted with the OptiLayer module OptiChar.

The crystalline structure of the single layer coatings was determined by grazing incidence X-ray diffraction (GIXRD). These measurements were performed using a PANalytical X’Pert PRO diffractometer with a Cu K$\alpha$ source. The incident angle was fixed at 1.5$^\circ{}$ and 2$\theta$ was scanned between 10$^\circ{}$ and 50$^\circ{}$.

Mechanical loss was measured using a ringdown technique, in which resonant modes of a fused silica disk sample are excited and the motion is allowed to freely decay. The disk was mounted by balancing it on a spherical lens (known as a gentle nodal suspension~\cite{Cesarini_2009}) which ensures minimal frictional losses for modes with nodes at the centre of the disk. The loss is calculated from the time constant of the exponential decay of the motion. The Young’s modulus and Poisson ratio of the material are estimated using a Bayesian approach from the mode frequency shifts measured experimentally which are fitted to a finite element model~\cite{vajente2020method}.

In addition to the loss measurements, the coating thermal noise (CTN) of the HR stack was measured directly \cite{Gras_2018}. The optical absorption of the HR stack was also measured using the high finesse cavity used for the thermal noise measurements, in which heating due to absorption changes the resonant frequency of the cavity. Measuring this frequency shift as a function of laser power, calibrated to the effect measured for a coating of known absorption, allows the absorption to be estimated. Total optical loss, the sum of absorption, scattering, and transmittance, can also be determined based on the cavity pole measurement. This allows scattering losses to be measured, since the transmittance of the HR stack was measured by spectrophotometry after annealing at 300$^\circ{}$C, a value that is assumed to not be modified by subsequent annealing steps.

\section{Results}

\subsection{TiO$_2$:GeO$_2$ single layer coatings}

\textit{Composition--} The composition obtained with RBS for TiO$_2$:GeO$_2$ single layers is presented in Table \ref{table:rbs}. The films were found to composed of Ti, Ge, O, and Ar as expected. The presence of Ar is due to the IBD deposition process during which the process gas is incorporated into the coatings. For all single layers grown in this work, the Ar content remained between 0.4 - 0.5 \% . In order to evaluate the TiO$_2$ doping proportion, the dopant cation ratio is presented and calculated as the ratio between the atomic concentration of Ti and the total (Ti+Ge) atomic concentration. Different dopant cation ratios were evaluated in this work, ranging from 0.37 to 0.52.

\begin{table}
    \centering
    \begin{tabular}{c|c|c|c|c|c|c} 
     \hline
    \multirow{2}{*}{Sample} & Ti & Ge & O & Ar & Dopant cation ratio & $n$ @ \\
    \cline{2-5}
    & \multicolumn{4}{c|}{(atomic \%)} & Ti/(Ti + Ge) & $\lambda$= 1064 nm\\
    \hline
    I & 12.3 & 20.6 & 67 & 0.4 & 0.374 & 1.83 $\pm$ 0.01\\
     II & 14.4 & 18.4 & 67 & 0.5 & 0.439 & 1.88 $\pm$ 0.01\\ 
    III & 15.1 & 16.5 & 68 & 0.4 & 0.478 & 1.91 $\pm$ 0.01\\ 
     IV & 16.4 & 15.1 & 68 & 0.5 & 0.521 & 1.95 $\pm$ 0.01\\ 
    \hline
\end{tabular}
    \caption{Composition of TiO$_2$:GeO$_2$ single layers obtained with RBS. The uncertainty for oxygen content is 2\%.}
    \label{table:rbs}
\end{table}

\begin{figure}[h!]
    \centering
    \includegraphics[width=0.85\columnwidth]{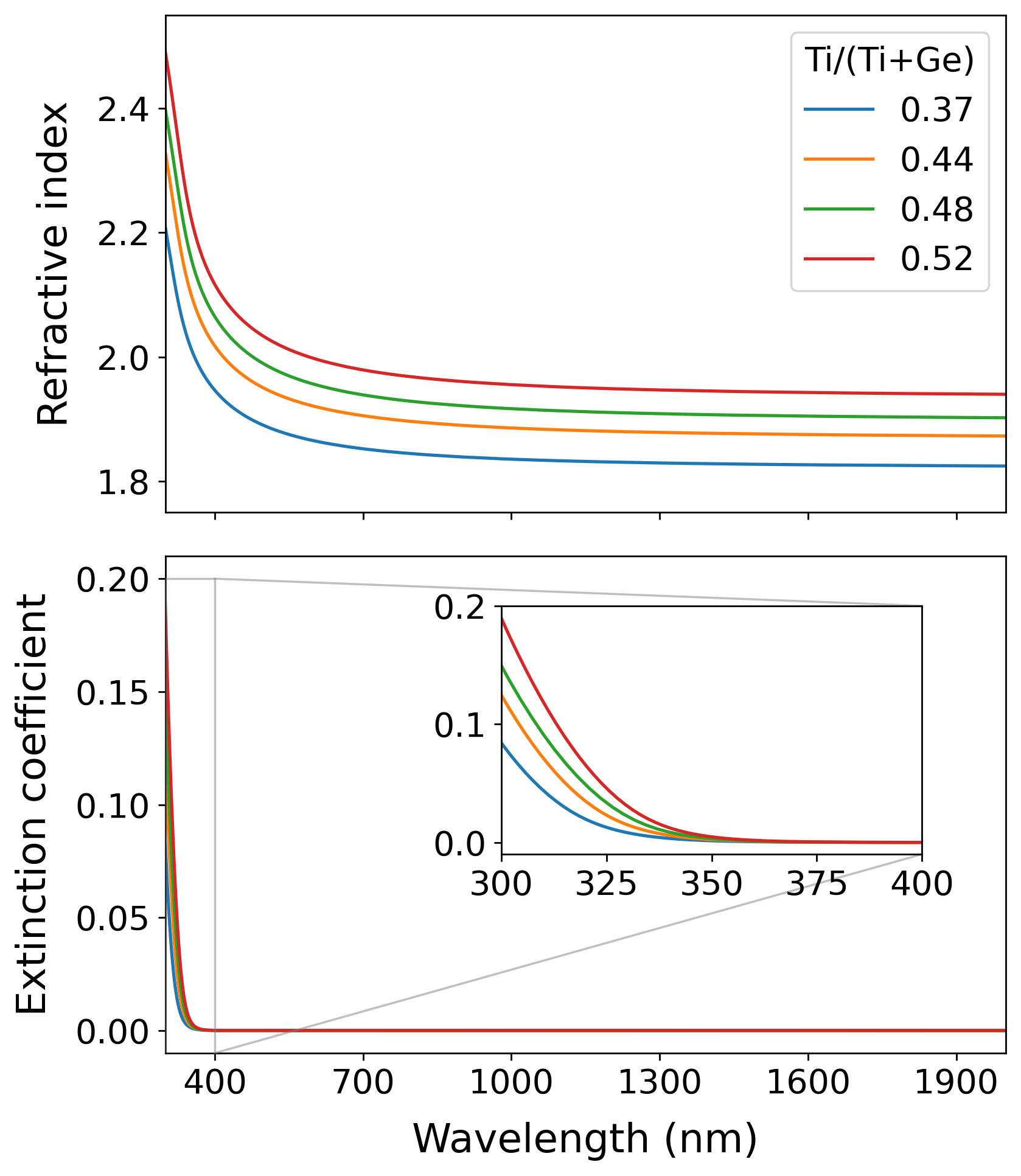}
    \caption{Refractive index (top panel) and extinction coefficient (bottom panel) of TiO$_2$:GeO$_2$ single layers. Inset shows the detail of the absorption tail at low wavelengths.}
    \label{fig:n-k}
\end{figure}

\textit{Optical properties---} Figure \ref{fig:n-k} shows the refractive index and extinction coefficient of TiO$_2$:GeO$_2$ single layers with different cation ratios. The refractive index at $\lambda$ = 1064 nm is listed in Table \ref{table:rbs} and shows an increase proportional to the increasing dopant cation ratio of the coatings. For the coating with a dopant cation ratio of 0.44, the refractive index at $\lambda$ = 1064 nm is in good agreement with previously reported values for IBD TiO$_2$:GeO$_2$ films with a dopant cation ratio of 0.446$\pm$0.003 \cite{vajente2021low}. The extinction coefficient falls rapidly to zero (within the sensitivity of this technique) with an absorption tail at low wavelengths that correlates to the dopant concentration in the coatings.

\begin{figure}[h!]
    \centering
    \includegraphics[width=0.85\columnwidth]{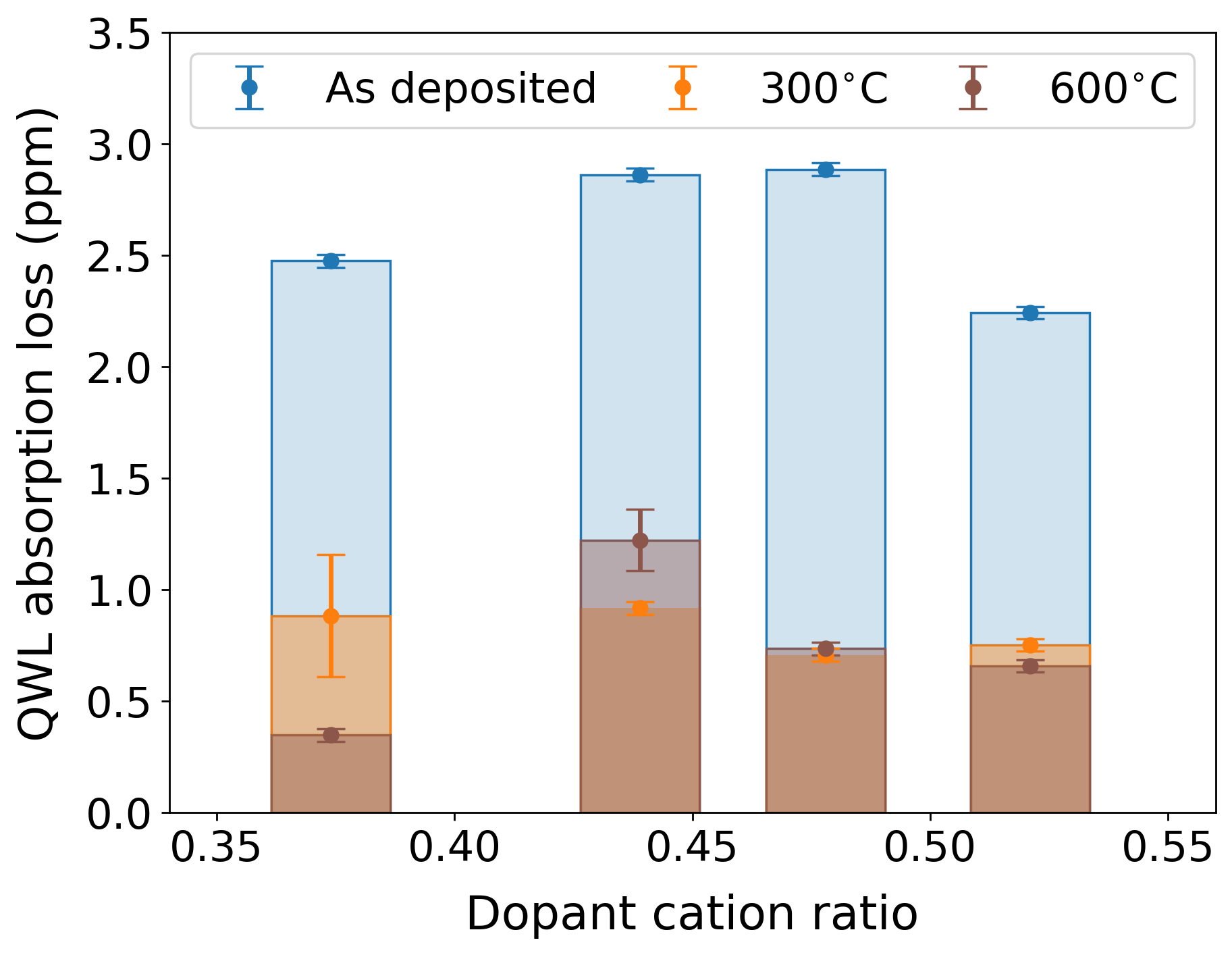}
    \caption{Optical absorption loss normalized to QWL optical thickness at $\lambda$ = 1064 nm for TiO$_2$:GeO$_2$ single layers as deposited and after annealing at 300$^{\circ}$C and 600$^{\circ}$C.}
    \label{fig:pci}
\end{figure}

\textit{Optical absorption---} For the TiO$_2$:GeO$_2$ single layers as deposited and after annealing, their optical absorption loss was measured by PCI at $\lambda$ = 1064 nm. The optical absorption loss was normalized to a quarter wavelength (QWL) optical thickness for ease of comparison between the different materials and is presented in Figure \ref{fig:pci}. In a typical HR stack composed of these coatings, each layer will have a QWL optical thickness thus the normalized absorption loss takes into account the variation in the refractive index due to the different dopant concentrations evaluated in this study. The QWL absorption loss for all the as-deposited coatings is between 2.5 - 3 ppm. After annealing at 300$^{\circ}$C, all coatings show a decrease in the absorption loss irrespective of the dopant concentration. However, after annealing at 600$^{\circ}$C most coatings show either a slight increase or no significant variation in the absorption loss except for the film with a dopant cation ratio of 0.37 for which the loss decreases significantly.

\begin{figure}
\begin{tabular}{cc}
  \includegraphics[width=0.5\columnwidth]{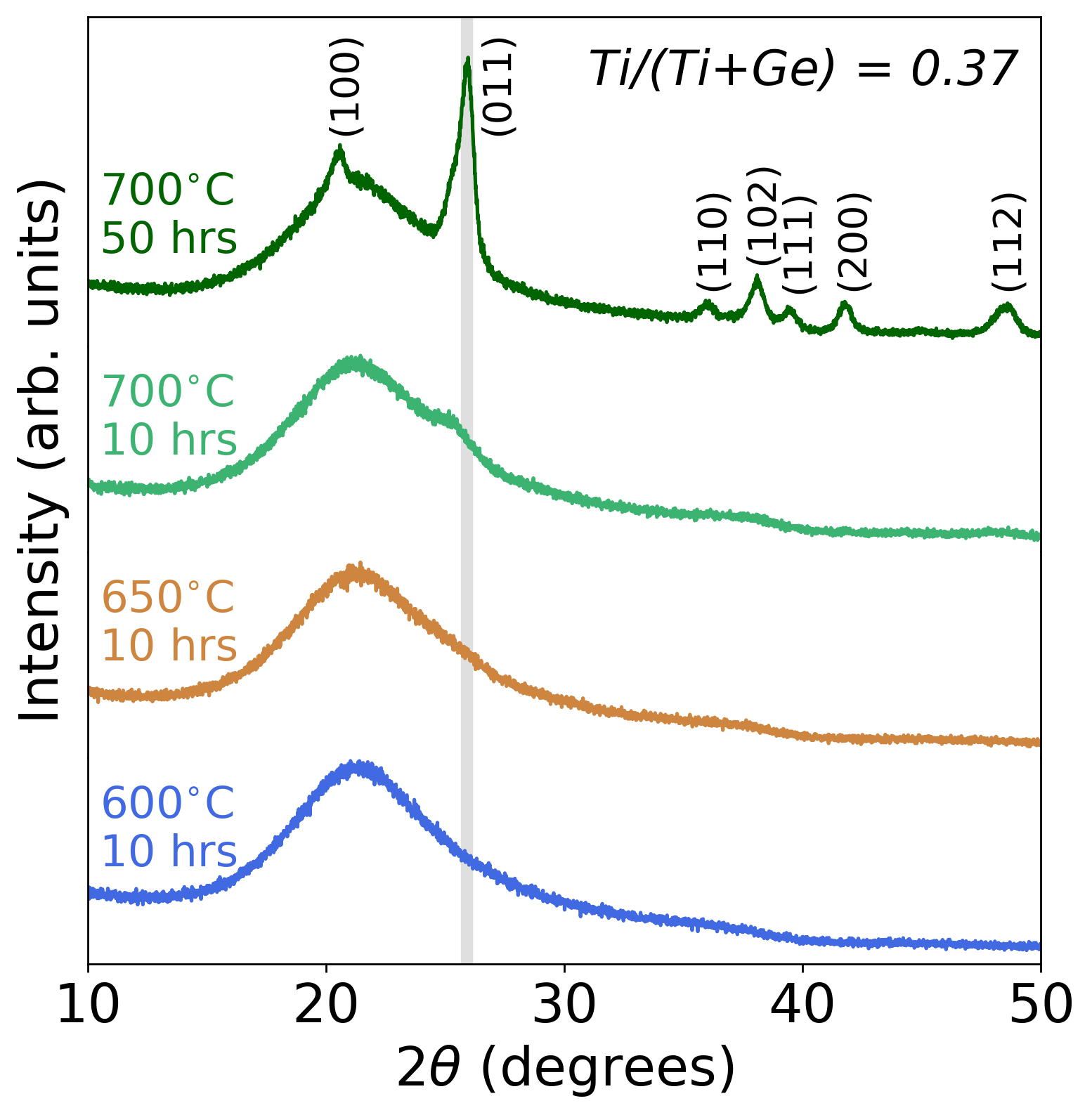} &   \includegraphics[width=0.5\columnwidth]{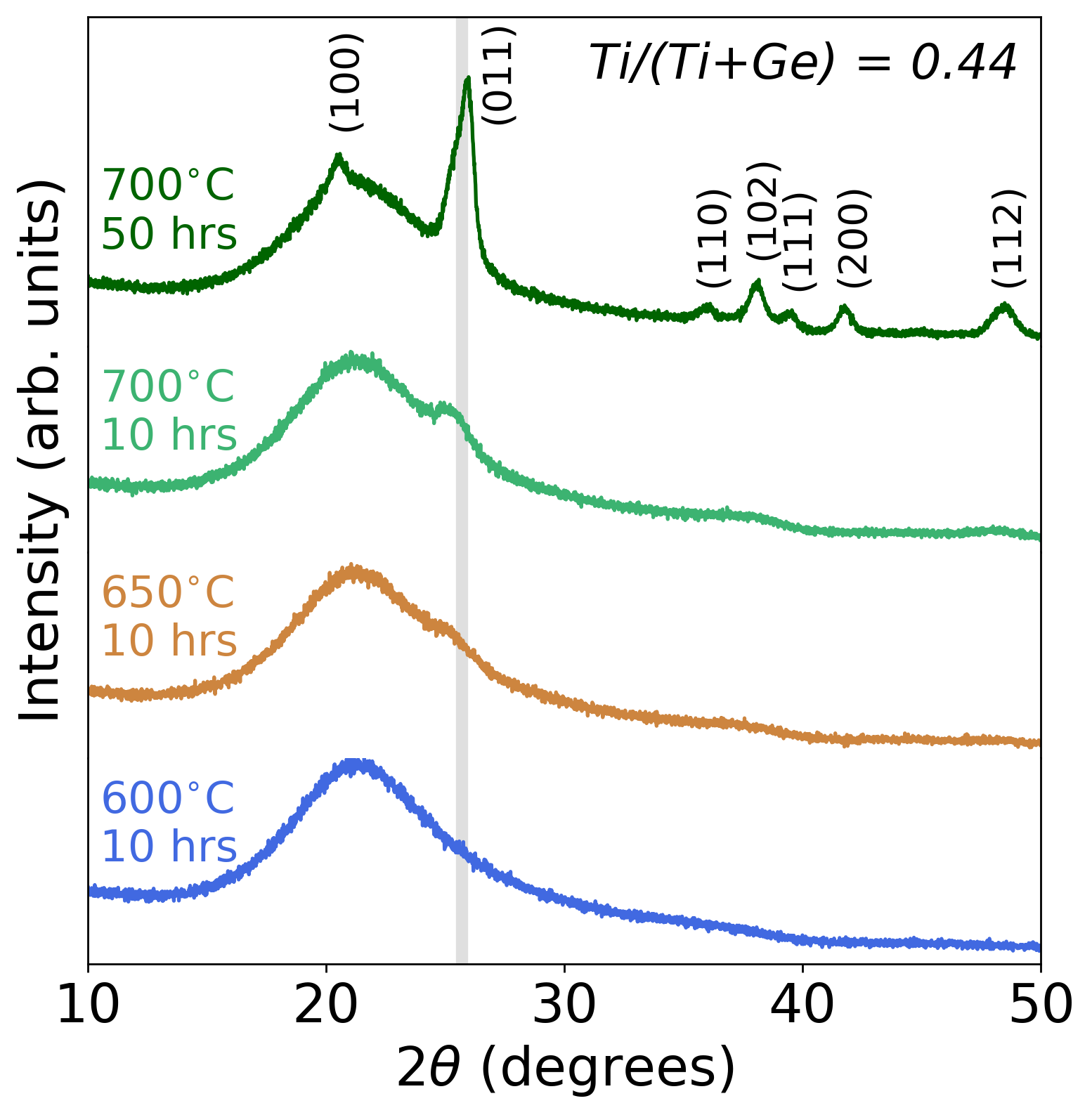} \\
 \includegraphics[width=0.5\columnwidth]{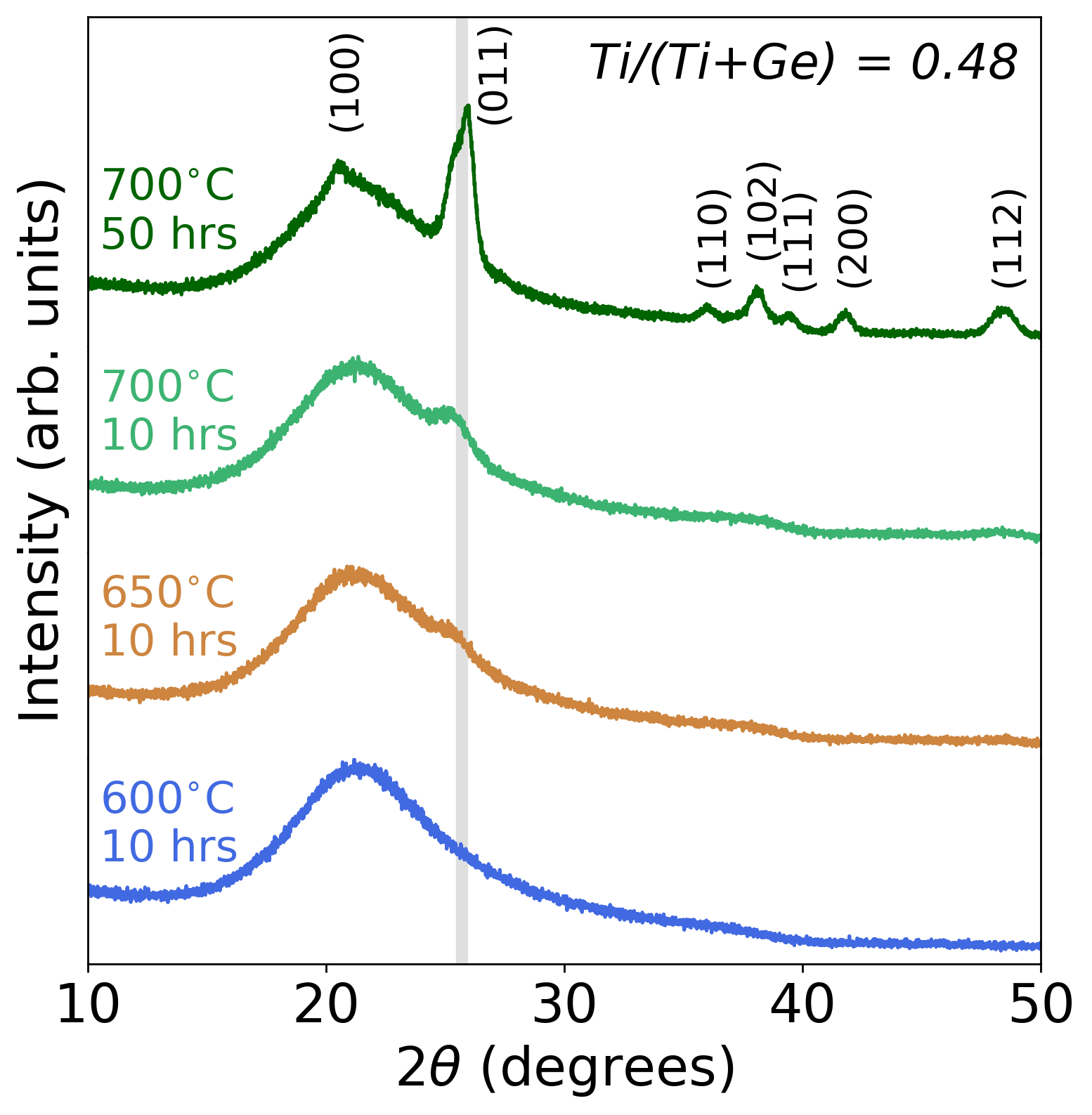} &   \includegraphics[width=0.5\columnwidth]{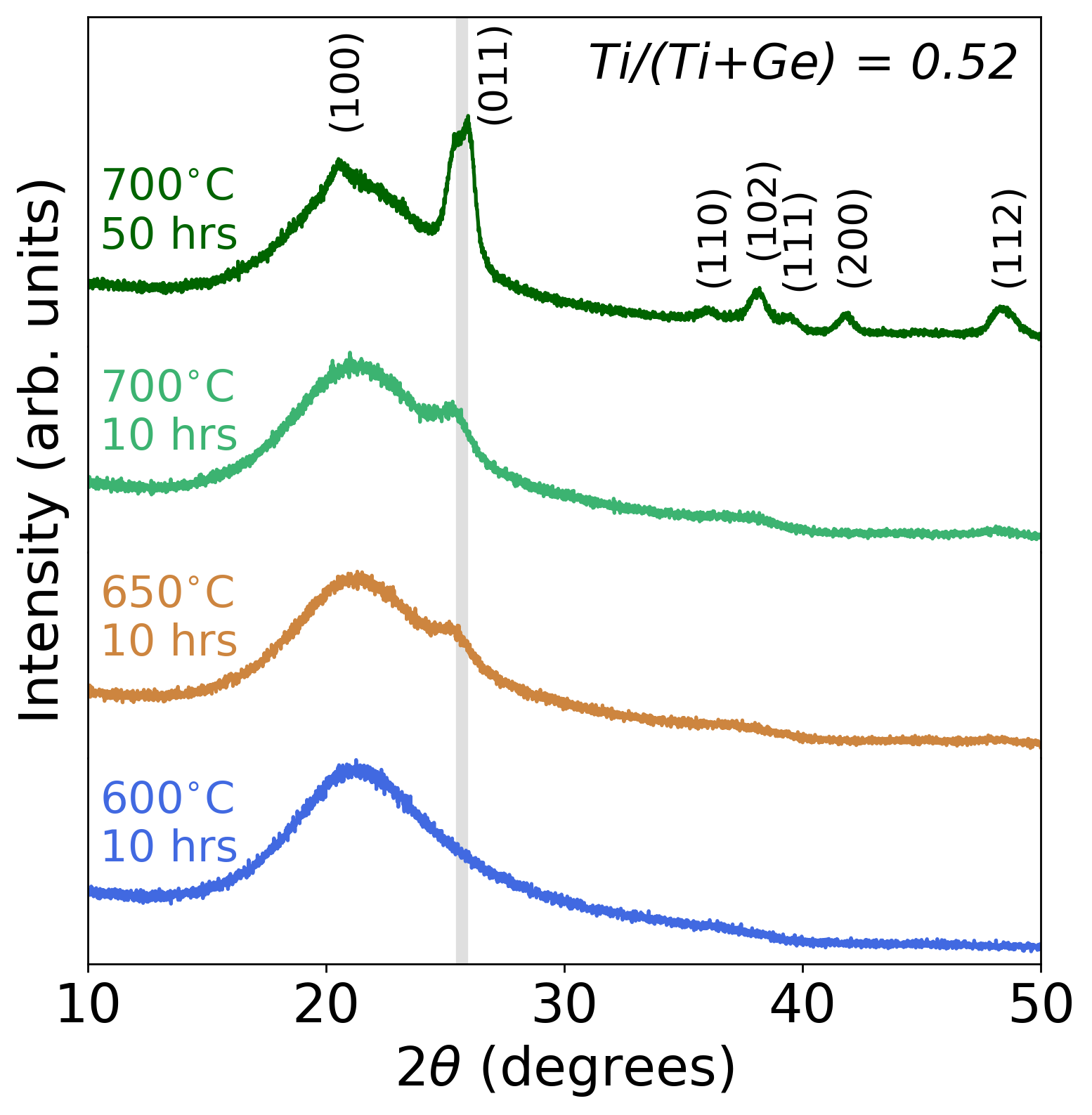} \\
\end{tabular}
\caption{GIXRD diffractograms of TiO$_2$:GeO$_2$ single layer coatings as deposited and after annealing at different temperatures. Tabulated peak positions for hexagonal GeO$_2$ (reference pattern PDF 01-083-0546~\cite{gates2019powder}) is included.}
\label{fig:gixrd}
\end{figure}

\textit{Crystalline structure--} The crystalline structure of the TiO$_2$:GeO$_2$ single layer coatings was evaluated with GIXRD after annealing under different conditions and diffractograms are shown in Figure \ref{fig:gixrd}. All films are amorphous as deposited and up to an annealing temperature of 600$^{\circ}$C, presenting only a broad feature around 2$\theta$ $\approx$ 22$^{\circ}$. For the lowest dopant cation ratio evaluated of 0.37, after annealing at 700$^{\circ}$C for 10 hrs a subtle peak can be identified at 2$\theta$ $\approx$ 26$^{\circ}$ and after further annealing for 50 hrs several sharp peaks can be observed consistent with partial crystallization of the coating. The structure was identified as hexagonal GeO$_2$ (reference pattern PDF 01-083-0546~\cite{gates2019powder}) and the peaks were indexed accordingly in Figure \ref{fig:gixrd}. The broad peak observed after annealing at 700$^{\circ}$C for 10 hrs can now be associated with GeO$_2$ (011). In the case of the rest of the evaluated coatings, evidence of the start of crystallization can already be identified after annealing at 650$^{\circ}$C for 10 hrs with the crystallized structure identified also as hexagonal GeO$_2$ (reference pattern PDF 01-083-0546~\cite{gates2019powder}). Previous studies have shown that GeO$_2$ coatings crystallize after annealing at 600$^{\circ}$C for 10 hrs~\cite{fazio2021comprehensive}. Doping GeO$_2$ with TiO$_2$ therefore increases the crystallization temperature and this is most effective at the lowest dopant concentration evaluated in this work. In addition, the TiO$_2$ content of the films show no signs of crystallization up until the maximum annealing temperature of 700$^{\circ}$C for 50 hrs regardless of the dopant concentration of the coatings.

\begin{figure}[h!]
    \centering
    \includegraphics[width=\columnwidth]{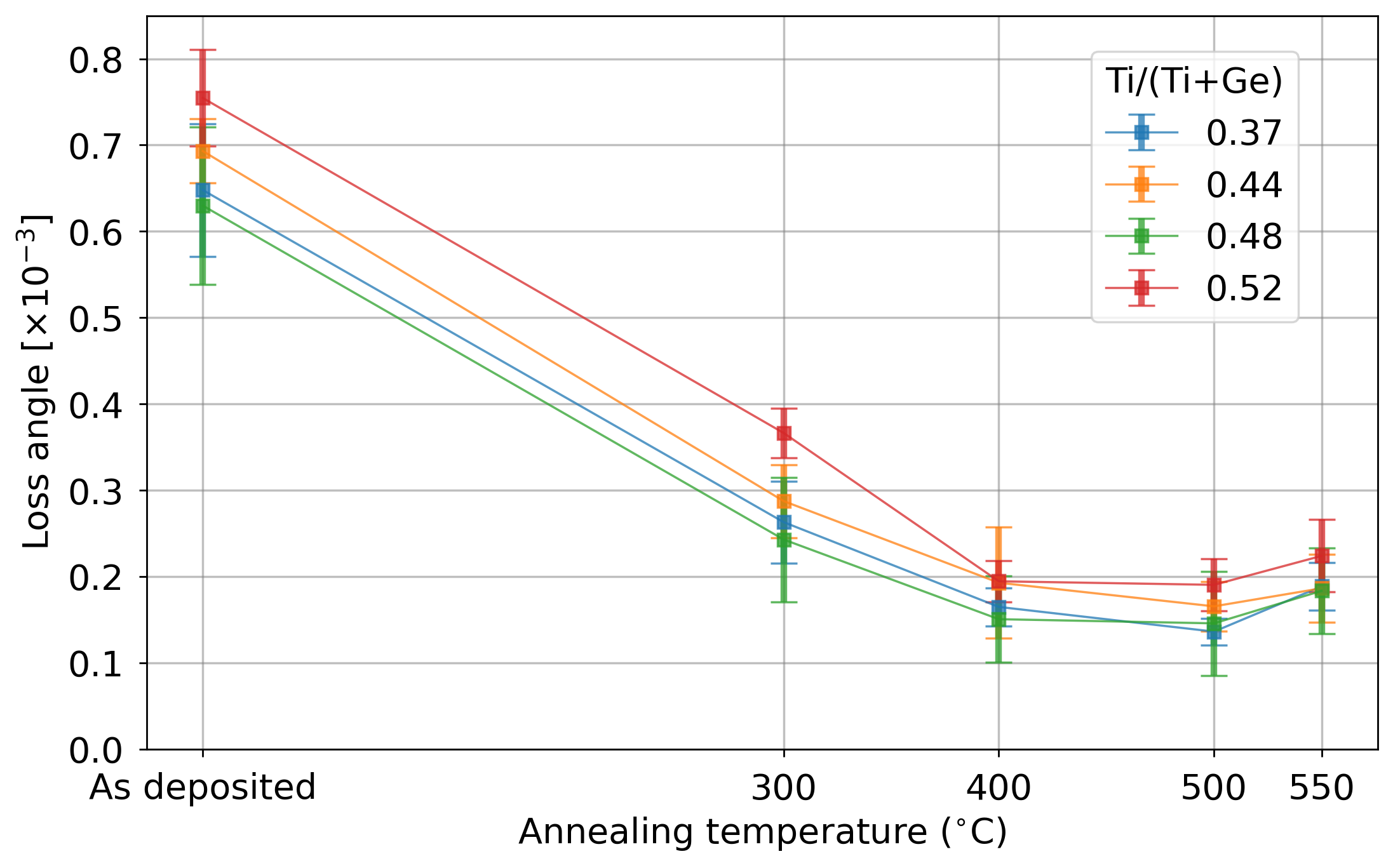}
    \caption{Coating loss angle at 1 kHz for titania-doped germania films with different dopant cation ratios as a function of the annealing temperature.}
    \label{fig:loss-angle}
\end{figure}

\textit{Mechanical loss--} The TiO$_2$:GeO$_2$ coating loss angle at 1 kHz as a function of annealing temperature is presented in Figure \ref{fig:loss-angle}. The as-deposited films feature loss angles in the range of 6 - 8 $\times$10$^{-4}$ which decrease with annealing irrespective of the dopant cation ratio. The decrease in the loss angle with annealing is slightly less pronounced for the film with a doping ratio of 0.52. After annealing at 550$^{\circ}$C, the loss angle for all coatings shows a subtle increase but this is not statistically significant. The lowest loss angle achieved for the coating with a dopant cation ratio of 0.48 is (1.5$\pm$0.9) $\times$10$^{-4}$ after annealing at 500$^{\circ}$C, which is in good agreement with previously reported values for this material \cite{vajente2021low}. The Young's modulus was determined to be (89$\pm$1) GPa and the Poisson ratio to be 0.25$\pm$0.05. The loss angle for pure GeO$_2$ coatings annealed at 500$^{\circ}$C was previously reported to be (1.00$\pm$0.14) $\times$10$^{-4}$ \cite{fazio2021comprehensive}, so the addition of the TiO$_2$ dopant does not significantly affect this value but in turn increases the refractive index making this mixture suitable to be employed as the high index layer in an HR stack.

\subsection{TiO$_2$:GeO$_2$/SiO$_2$ HR multilayer stack}

\begin{figure}[h!]
    \centering
    \includegraphics[width=0.9\columnwidth]{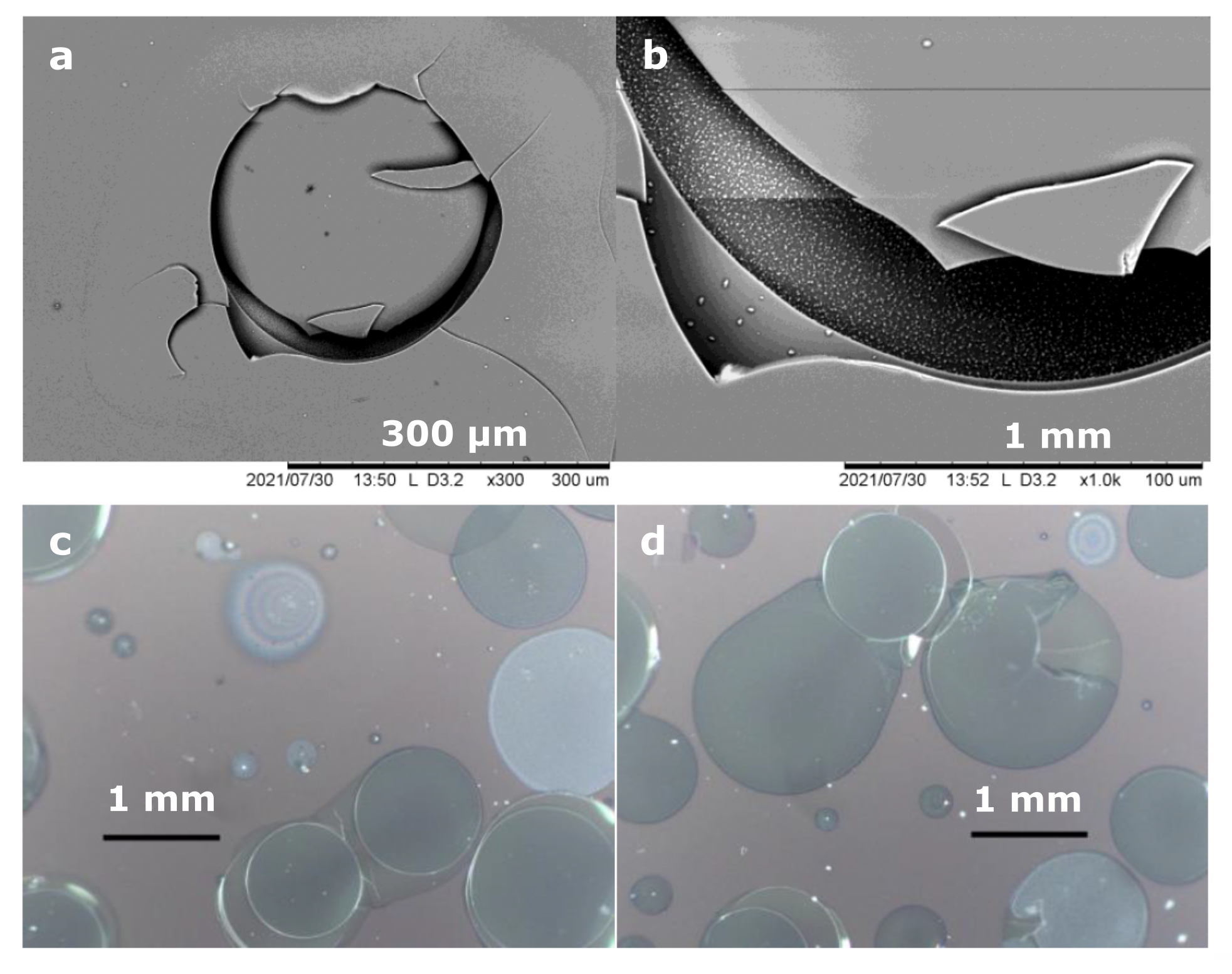}
    \caption{SEM (top) and optical microscope (bottom) images of the surface of a TiO$_2$:GeO$_2$/SiO$_2$ HR multilayer stack after annealing at 600$^{\circ}$C for 100 hrs.}
    \label{fig:blisters}
\end{figure}

\textit{Annealing induced defects--} Annealing creates localized defects in TiO$_2$:GeO$_2$/SiO$_2$ HR multilayer stack. After annealing at 300$^{\circ}$C circular blisters appear which increase in size and quantity until the last annealing step at 600$^{\circ}$C for 10 hrs. Extended annealing at that temperature for 100 hrs does not seem to modify the defects already present in the sample. Figure \ref{fig:blisters} shows the sample surface after the extended annealing. The top scanning electron microscope (SEM) images show localized delamination damage and cracks that extend from the blisters outwards. The optical microscope images indicate these blisters have varying sizes, ranging from hundreds of $\mu$m upwards, and appear to originate at different depths inside the stack.
This is in good agreement with reports of blisters in a similar TiO$_2$:GeO$_2$/SiO$_2$ HR multilayer stack by Rezac \textit{et. al.}~\cite{rezac2023imaging} and Lalande \textit{et. al.}~\cite{lalande2024ar}. Rezac \textit{et. al.} observed these defects using an air annealing scatterometer that showed the onset of small blisters after annealing at 500$^{\circ}$C with subsequent higher annealing temperatures leading to blister growth and coalescence. Lalande \textit{et. al.} reported that blister formation has been mitigated mainly by reducing the water partial pressure in the deposition chamber. The authors also found that Ar accumulation is a driver for blister growth. Fazio \textit{et. al.} have previously reported the formation of nanometer sized Ar-filled voids after annealing in TiO$_2$:Ta$_2$O$_5$ single layers grown by IBD \cite{fazio2020structure}. However, the role of Ar in blisters formation and evolution in IBD coatings is not entirely understood.

\begin{figure}[h!]
    \centering
    \includegraphics[width=\columnwidth]{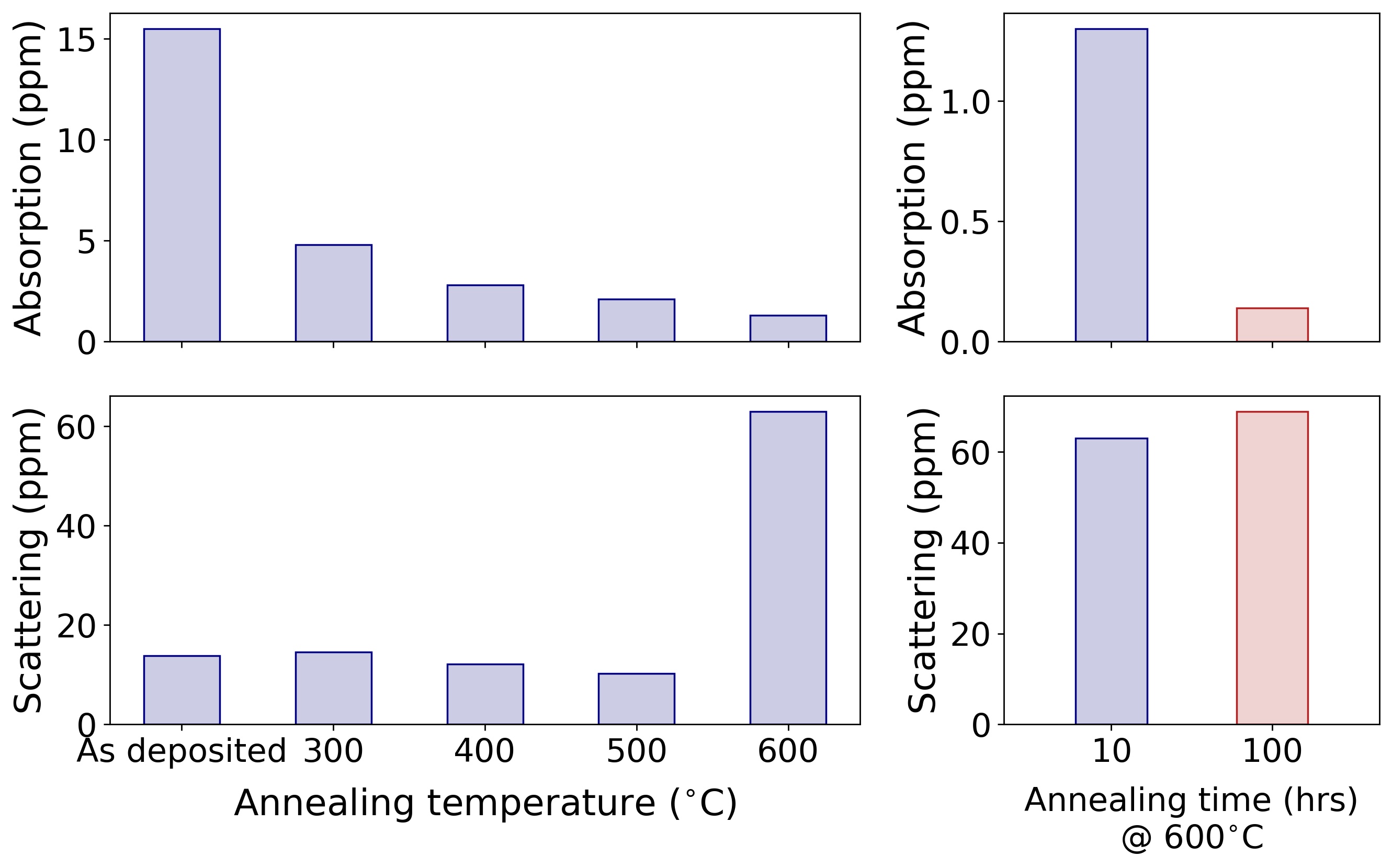}
    \caption{(a) Absorption loss and (b) scattering loss for an HR multilayer stack of TiO$_2$:GeO$_2$/SiO$_2$.}
    \label{fig:losses-hr}
\end{figure}

\textit{Absorption and scattering losses--} Despite the visible defects on the TiO$_2$:GeO$_2$/SiO$_2$ HR surface, an area near the centre of the sample was found to be defect free and measurements of absorption and total optical loss (from which scattering loss is derived) were performed there in the CTN optical cavity. The absorption and scattering losses of the TiO$_2$:GeO$_2$/SiO$_2$ HR multilayer stack were evaluated after annealing as shown in figure \ref{fig:losses-hr}. Annealing at increasing temperatures results in the reduction of optical absorption losses from 15.5 ppm as deposited to 1.3 ppm at 600$^{\circ}$C. Further annealing at 600$^{\circ}$C for an additional 100 hrs decreases the losses to 0.14 ppm, well below 0.5 ppm which is the required absorption value for current gravitational-wave detectors and in particular for A+ LIGO.

Scattering losses also decrease with annealing, however, there is a sharp increase after annealing at 600$^{\circ}$C. Upon extended annealing at 600$^{\circ}$C for 100 hrs, a slight increase in scattering is observed of about 10\%. This increase in scattering losses could be due to the increase of blisters causing surface damage that extends to the measured central area of the HR stack. Another possibility could be the onset of partial crystallization, which is detected by GIXRD at 650$^{\circ}$C however scattering losses might be more sensitive to the presence of few crystallites in the sample since the penetration depth is larger.

\begin{figure}[h!]
    \centering
    \includegraphics[width=\columnwidth]{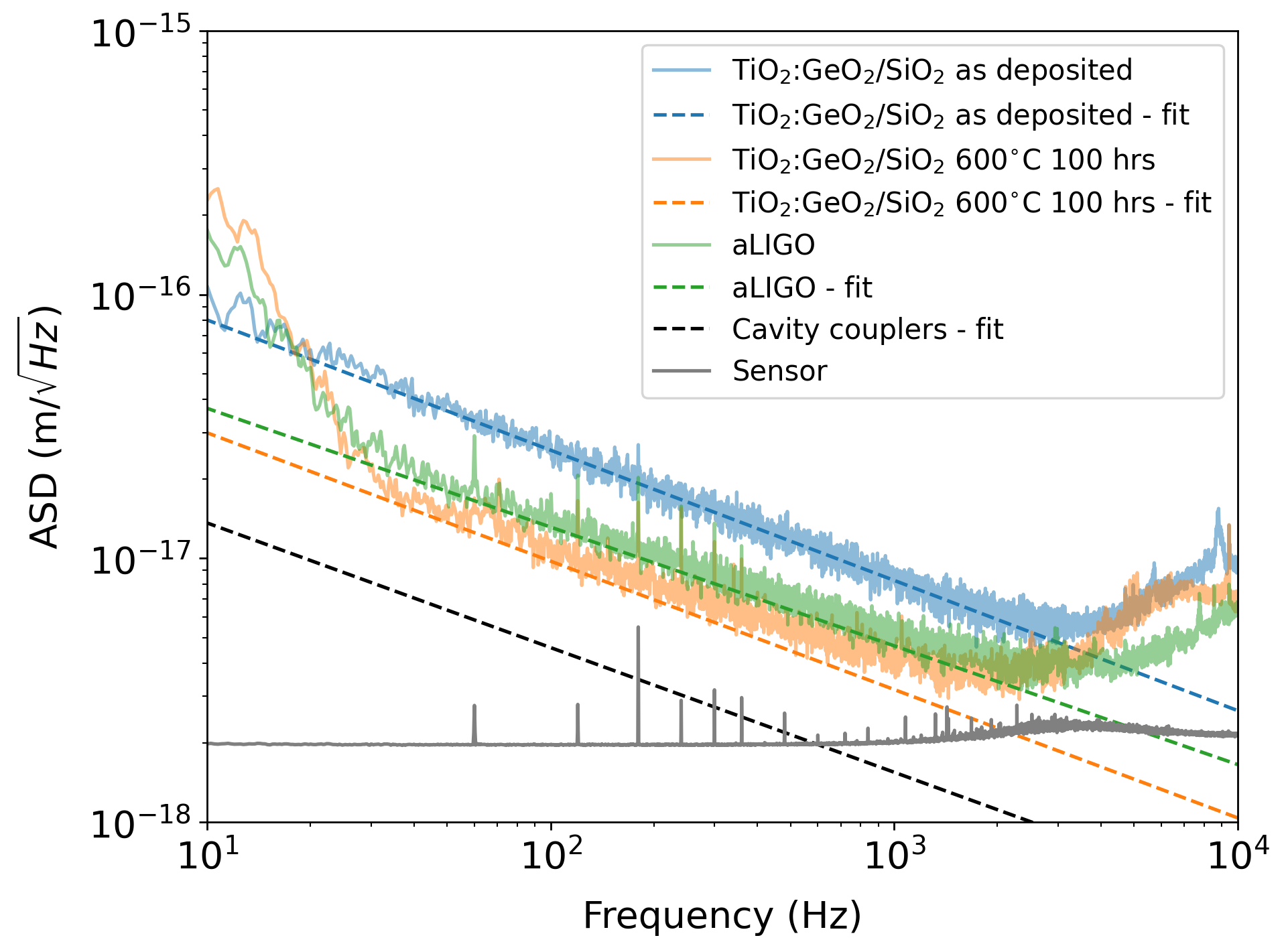}
    \caption{Amplitude spectral density (ASD) for an HR multilayer stack of TiO$_2$:GeO$_2$/SiO$_2$ as deposited and annealed at 600$^{\circ}$C for 100 hrs. For comparison, the Advanced LIGO noise spectrum measured from ETM witness samples is also shown. The sensor and cavity couplers noise are employed in the data fit as the measurements are a combination of CTN and these noise sources.}
    \label{fig:ctn}
\end{figure}

\textit{Coating thermal noise--} The CTN of the TiO$_2$:GeO$_2$/SiO$_2$ HR multilayer stack as deposited and after annealing at 600$^{\circ}$C for 100 hrs is presented in figure \ref{fig:ctn}. The measured data and fit are shown, along with the noise spectrum of an Advanced LIGO witness sample measured in the same apparatus for comparison. For the annealed coating, the central area that appeared free of defects was measured. The as-deposited HR stack shows a higher CTN than the ETM witness sample from Advanced LIGO, but after each annealing step the CTN is reduced. After annealing at 500$^{\circ}$C for 10 hrs (not pictured) the CTN of the stack reaches the value of the Advanced LIGO ETM witness sample and is further reduced to about 74\% of that of Advanced LIGO at 100 Hz after annealing at 600$^{\circ}$C for 100 hrs. This reduction in CTN is expected from the measured decrease in the mechanical loss of single layers, however, the reduction in loss is smaller than predicted by Vajente \textit{et. al.}~\cite{vajente2021low}. Since in this study single layers were not annealed for 100 hrs for mechanical loss measurements, we will consider the loss values after annealing at 500$^{\circ}$C for 10 hrs from figure \ref{fig:loss-angle}. Using the same effective medium approach outlined by Vajente \textit{et. al.}~\cite{vajente2021low}, the calculated CTN at 500$^{\circ}$C is nearly 90\% of that of Advanced LIGO whereas the measured value is about the same as the Advanced LIGO ETM witness sample. This indicates that there is excess CTN in the HR stack compared to the estimate from single layers, which could be due to interfacial effects, differences between the bulk and shear mechanical losses, or other phenomena affecting the stack.

\section{\label{sec:conclusions} Conclusions}

We presented the optical and structural characterisation of TiO$_2$:GeO$_2$ single layers with dopant cation ratios ranging from 0.374 - 0.521 as deposited and after heat treatment at different temperatures. We show the optical constants for each material, which follow the expected law of mixtures, and found that the coatings feature partial crystallization of the GeO$_2$ phase after annealing at 650$^{\circ}$C except from the film with the lowest cation ratio. The absorption loss and mechanical loss in all coatings is shown to decrease after annealing. The loss angle reaches values as low as 1 $\times$10$^{-4}$ in agreement with previous reports \cite{vajente2021low}. Based on this, a TiO$_2$:GeO$_2$/SiO$_2$ HR stack was designed and deposited. This HR stack is demonstrated to feature a coating thermal noise reduction of 25\% compared to the current end test masses of Advanced LIGO. The observed blistering found when annealing the HR stack lead to increased scattering losses, which could be reduced by limiting the water partial pressure in the IBD chamber. Most notably, the exceptionally low absorption loss (sub-ppm) well below the requirements of A+ LIGO, combined with the reduction in thermal noise, indicates that TiO$_2$:GeO$_2$ is a material candidate for use in gravitational-wave detectors that can lead to an increased sensitivity and astronomical reach.

\begin{acknowledgments}
We are grateful for financial support from STFC (ST/V005642/1, ST/V005634/1, ST/V005626/1) and the University of Strathclyde. M. F. is supported by an STFC Ernest Rutherford Fellowship (ST/W004844/1) and S.R. is supported by a Royal Society Industry Fellowship. Work done at U. Montreal is supported by the NSERC and the FRQNT through the RQMP. We are grateful to the National Manufacturing Institute for Scotland (NMIS) for support, and we thank our colleagues in the LSC and Virgo collaborations and within SUPA for their interest in this work. Work done at Cal State Fullerton was supported by NSF grant 2207998. This paper has LIGO Document No. P2300359.
\end{acknowledgments}

\bibliography{biblio}

\end{document}